\documentclass[aps,prb, superscriptaddress,twocolumn,showpacs,longbibliography]{revtex4-2}

\usepackage[utf8]{inputenc}
\usepackage{amsmath,amssymb}
\usepackage{graphicx}
\usepackage[pdfencoding=auto,psdextra]{hyperref}
\usepackage{color}
\usepackage{float}
\usepackage{caption}
\usepackage{subcaption}

\usepackage{placeins}
\usepackage{dblfloatfix} 
\hypersetup{
    colorlinks=true,
    linkcolor=blue,
    citecolor=blue,
    urlcolor=blue
}

\DeclareCaptionLabelFormat{fig}{Fig.~#2}
\begin{document}

\title{Geometric dependence of exchange bias in tilted three-dimensional CoFe/IrMn microwires}

\author{Balram Singh}
\email{balram.singh@tuwien.ac.at}
\affiliation{Institute of Applied Physics, TU Wien, 1040 Vienna, Austria}

\author{Aman Singh}
\affiliation{Leibniz Institute for Solid State and Materials Research Dresden, 01069 Dresden, Germany}
\affiliation{Institute of Applied Physics, TU Dresden, 01187 Dresden, Germany}

\author{Stefan Mikulik}
\affiliation{Institute of Applied Physics, TU Wien, 1040 Vienna, Austria}

\author{Jakub Jurczyk}
\affiliation{Institute of Applied Physics, TU Wien, 1040 Vienna, Austria}

\author{Volker Neu}
\affiliation{Leibniz Institute for Solid State and Materials Research Dresden, 01069 Dresden, Germany}

\author{Amalio Fernández-Pacheco}
\email{amalio.fernandez-pacheco@tuwien.ac.at}
\affiliation{Institute of Applied Physics, TU Wien, 1040 Vienna, Austria}

\date{\today}

\begin{abstract} The exchange bias (EB) effect, arising from interfacial coupling between ferromagnetic (FM) and antiferromagnetic (AF) layers, induces a unidirectional magnetic anisotropy and underpins a wide range of spintronic functionalities. Extending the EB effect to three-dimensional (3D) architectures enables investigation of interfacial coupling in non-planar structures, which is a key step toward realizing spintronic functionalities beyond planar systems. Achieving this requires the fabrication of FM/AF bilayers with smooth interfaces and well-defined thicknesses on non-planar scaffolds, together with suitable characterization methods.
In this work, we realize exchange-biased 3D FM/AF microwires by combining two-photon lithography with magnetron sputtering. CoFe/IrMn bilayers are deposited on microwire scaffolds with inclination angles of $0^\circ$, $30^\circ$ and $45^\circ$ relative to the substrate, and their magnetization reversal is probed using dark-field magneto-optical Kerr effect (DF-MOKE) magnetometry. We find that the EB and coercive fields vary in a characteristic way with the inclination angle, consistent with the systematic reduction in film thickness expected from inclined directional deposition. In addition, the EB magnitude is influenced by the combined effects of surface roughness of non-planar geometries and the directional growth of the bilayer, highlighting the importance of 3D scaffold surface quality for integrating magnetic multilayers. These results provide insight into the growth and magnetic behavior of sputter-deposited magnetic multilayers with functional interfaces on 3D geometries.

\end{abstract}

\maketitle


\section{Introduction} The extension of nanomagnetism into three-dimensional (3D) architectures has opened new avenues for tailoring magnetic functionalities through geometric parameters such as thickness gradients, torsion and curvature.~\cite{Gubbiotti2025Roadmap3D, MakarovSheka2022, FernandezPacheco2017} These additional degrees of freedom enrich conventional magnetic phenomena and enable the emergence of effects absent in planar systems, including complex spin textures,~\cite{Volkov2019, Sheka2020, Tretiakov2017, Singh2025, Xu2025} modified magnetization dynamics~\cite{Hertel2013, Skoric2020, Yan2011, Otalora2016} and geometry-induced magnetic interactions.~\cite{Guo2023, Sahoo2021} As a result, 3D nanomagnetic architectures hold great potential for next-generation technologies,~\cite{Dieny2020} such as ultrahigh-density data storage,~\cite{Gu2024, Fedorov2024, Parkin2008} logic devices,~\cite{Becherer2021} sensors,~\cite{Karnaushenko2015GMI, Rivkin2021} interconnects~\cite{Burks2020} and neuromorphic computing.~\cite{Dion2024} Realizing such complex 3D magnetic architectures requires coordinated advances in fabrication and characterization, particularly since the growth of magnetic multilayers with high-quality interfaces on non-planar geometries remains a persistent challenge.

Within this context, exchange bias (EB) is a well-established interfacial magnetic phenomenon arising from exchange coupling between a ferromagnet (FM) and an antiferromagnet (AF).~\cite{Meiklejohn1956} This interaction induces a unidirectional magnetic anisotropy in the FM layer and plays a crucial role in a wide range of spintronic devices, including spin valves and non-volatile magnetic memories.~\cite{Cowburn2003} Extensive experimental and theoretical studies~\cite{Blachowicz2021, Stamps2000, Berkowitz1999} have shown that both the magnitude and stability of EB are governed by factors such as the FM/AF interface quality,~\cite{Vafaee2016, Nogues1999} the properties of the AF layer,~\cite{Morales2009} and the thickness and spin configuration of the FM layer.~\cite{Yu2000, Leighton2002, Morales2015} In particular, the EB field ($H_{\mathrm{EB}}$) is shown to scale inversely with the FM thickness $t^{\mathrm{FM}}$, i.e., $H_{\mathrm{EB}} \propto 1/t^{\mathrm{FM}}$, a result of the interfacial origin of this effect.~\cite{Leighton2002, Nogues1999, Yu2000}

Building on the understanding of EB in planar systems and given its strong sensitivity to interfacial quality and thickness variations, extending EB systems to 3D geometries enables us to investigate the impact of non-planar geometries on bilayer growth, interface quality and thickness gradients.

\captionsetup[figure]{labelfont=bf}
\begin{figure*}[ht!]
    \includegraphics[width=1.0\linewidth]{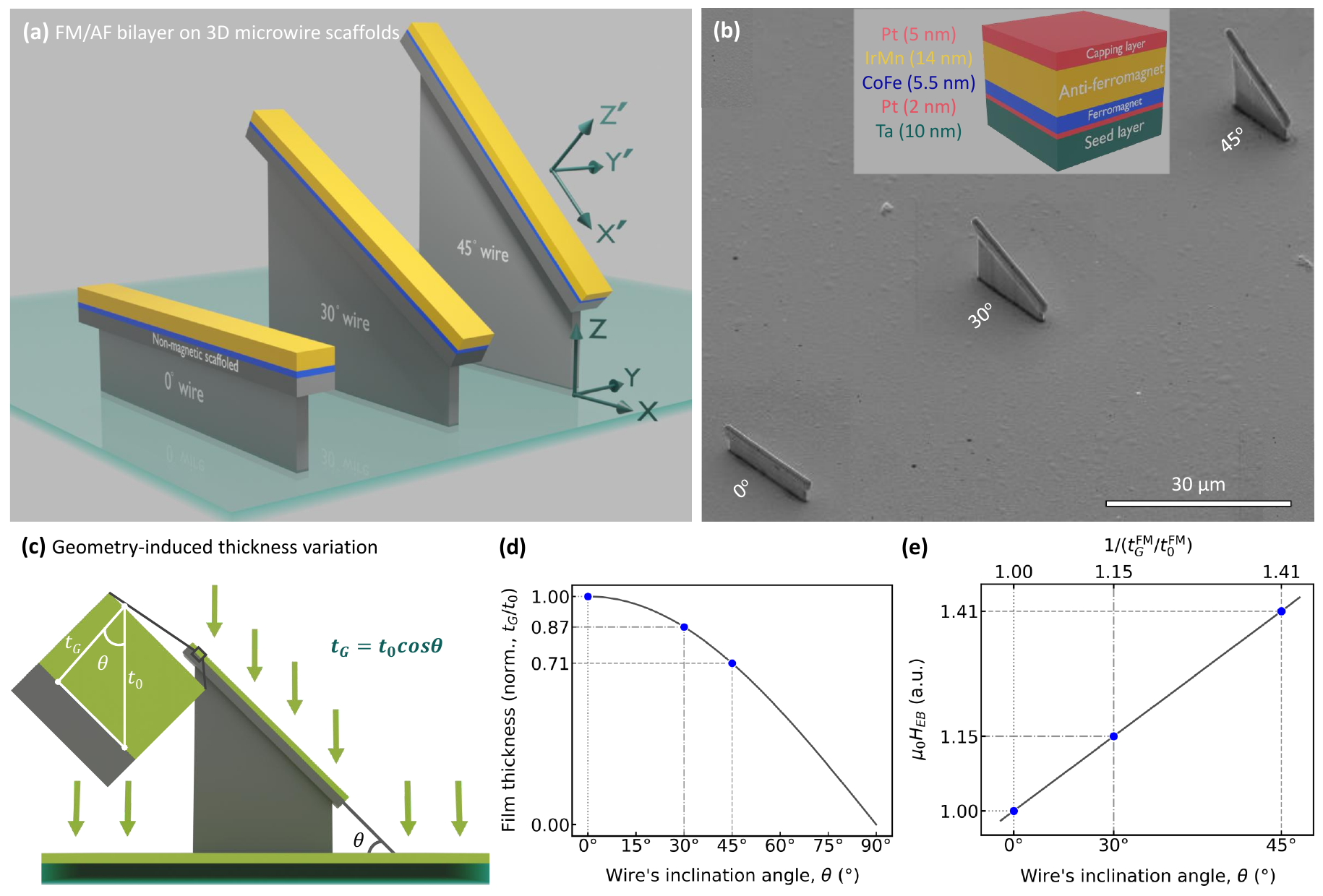}
    \caption{\textbf{Overview of the exchange-biased 3D CoFe/IrMn microwires.}
\textbf{(a)} Schematic of the 3D wires with inclination angles of 0°, 30° and 45° coated with a FM/AF bilayer. Note the two coordinate systems: global ($x$-$y$-$z$) and local ($x'$-$y'$-$z'$).
\textbf{(b)} Scanning electron microscopy (SEM) images of the 3D CoFe/IrMn microwires (length 20~$\mu$m and width 1~$\mu$m) of inclination angles of 0°, 30° and 45°. The inset presents the deposited layer stack on the 3D wire scaffolds.
\textbf{(c)} Schematic illustration of the sputter-deposition geometry with respect to the wire inclined at an angle $\theta$, highlighting the geometric thickness of the layer, $t_{G}$, compared to the nominal thickness on the substrate $t_{0}$.
\textbf{(d)} Expected reduction in the layers (FM and AF) thickness as a function of inclination angle, $\theta$, for a highly directional deposition process.
\textbf{(e)} Illustration of expected increase in the $H_{EB}$ with inclination angle $\theta$, arising due to the reduced geometric FM thickness $t_{G}^{FM}$, based on purely geometrical deposition arguments. The top horizontal axis represents the inverse of the normalized $t_{G}^{\mathrm{FM}}$.}
    \label{fig:main_figure}
\end{figure*}

Among the available methods for fabricating 3D nonmagnetic micro- and nano-structures, ~\cite{MakarovSheka2022, FernandezPacheco2017, Meng2021, Meng2021ACS} two-photon lithography (TPL) offers exceptional geometrical freedom with submicron precision, smooth surfaces and high fabrication throughput,~\cite{Askey2025, GuMetalens3DNanolithography} making it suitable for fabricating complex 3D polymeric scaffolds. These scaffolds can subsequently be coated with magnetic layers using techniques such as electrodeposition,~\cite{Askey2020} atomic layer deposition (ALD),~\cite{Xu2025} or physical vapor deposition (PVD)~\cite{Askey2024DomainWallPinning3D} to achieve 3D functional magnetic micro- and nano-structures. While ALD and electrodeposition provide conformal coatings, they are often limited by a low number of material choices and the quality of the interfaces in multilayers.~\cite{Stano2018a} In contrast, PVD is widely used for spintronic multilayer stacks due to its ability to fabricate multilayers with low surface roughness and high-quality interfaces. However, its directional nature can lead to thickness gradients and modified film microstructure when applied to non-planar scaffolds.~\cite{Hawkeye2014, Lintymer2003, Skoric2020} To overcome these limitations, strategies such as growing multilayers on planar substrates and then transforming them into the 3D shapes via stress engineering~\cite{Schmidt2001, Streubel2016} or transferring multilayers onto non-planar substrates,~\cite{Gu2024} have been explored. However, these approaches may potentially modify the magnetic layer properties, for instance, due to the strain induced during reshaping and are also limited to the achievable geometries.

In this work, we combine TPL and PVD to realize exchange-biased 3D FM/AF systems (Fig.~\ref{fig:main_figure}a,b) and systematically investigate the impact of non-planar geometry on bilayer growth, specifically film thickness, microstructure and interface quality, as well as the resulting EB behavior. We fabricate 3D wire scaffolds with inclination angles of $\theta = 0^\circ$, $30^\circ$ and $45^\circ$ relative to the substrate, each with a constant length of 20 $\mu$m and width of 1 $\mu$m. The scaffolds are coated with FM/AF bilayers using magnetron sputtering. The investigated sample geometries include: (i) a continuous FM/AF bilayer deposited on a flat substrate, hereafter referred to as a ``FM/AF planar film" and (ii) FM/AF bilayers deposited on wire scaffolds inclined by an angle $\theta$, hereafter referred to as ``FM/AF $\theta^\circ$ wires".

Considering the directional nature of sputter deposition, from the geometric arguments, the film thickness on inclined wires is expected to scale with the cosine of the inclination angle (Fig.~\ref{fig:main_figure}c,d). Therefore, for a nominal deposited thickness $t_0$, the thickness on the $\theta^\circ$ wire, here referred to as geometric thickness, is given by $t_G = t_0 \cos\theta$. Assuming comparable interface quality across different inclinations and a predominantly geometric thickness reduction, the EB field is therefore expected to scale as $H_{\mathrm{EB}} \propto 1/t_G^{\mathrm{FM}}$, Fig.~\ref{fig:main_figure}e.

To further validate this scaling, the nominal FM and AF layer thicknesses are varied between deposition runs, resulting in five exchange-biased FM/AF systems. Three samples form an FM-thickness series (CoFe5.5/IrMn14, CoFe7/IrMn14 and CoFe8.5/IrMn14), while three samples constitute an AF-thickness series (CoFe5.5/IrMn14, CoFe5.5/IrMn5 and CoFe5.5/IrMn3). The CoFe5.5/IrMn14 bilayer serves as a common reference for both series. The numbers indicate the thicknesses of the layers in nanometers (nm).The scaling of the $H_{\mathrm{EB}}$ with the nominal FM thickness, $t_0^{\mathrm{FM}}$, is then compared to its scaling with the geometric FM thickness, $t_G^{\mathrm{FM}}$, within a single deposition. This comparison enables us to evaluate whether the EB dependence in 3D wires can be fully understood in terms of arguments related to the deposition geometry.

Characterizing the magnetic properties of 3D architectures remains challenging, as most established techniques are optimized for planar thin films. Conventional magneto-optical imaging can probe only limited regions of structures with small curvature and loses sensitivity for large curvature or steeply inclined geometries.~\cite{Singh2022a, Singh2022b} X-ray-based magnetic imaging techniques~\cite{Streubel2015CurvedFilms} provide full three-dimensional sensitivity across a wide range of curvatures, materials and length scales, but require complex sample preparation and access to synchrotron radiation facilities. Magnetotransport measurements~\cite{Singh2022a, Singh2025} offer a comparatively accessible and indirect probe of magnetic anisotropy; however, they necessitate electrical contacting of 3D structures, which remains a significant fabrication challenge.

Here, we employ a custom-built dark-field magneto-optical Kerr effect (DF-MOKE) setup \cite{SanzHernandez2017} capable of applying magnetic fields in arbitrary directions.~\cite{CascalesSandoval2025} This technique offers high sensitivity for measuring magnetic moments \cite{SanzHernandez2023JAP} from individual 3D micro- and nanostructures with inclination angles ranging from $0^\circ$ to $50^\circ$, which motivates the specific design of the investigated 3D geometries. 

In this work, we investigate EB in 3D microwires with varying inclinations, focusing on how the directional nature of sputter deposition affects the effective FM and AF thicknesses, as well as the interfacial properties of CoFe/IrMn bilayers. By comparing smooth and rougher 3D scaffolds, together with planar bilayers of systematically varied layer thicknesses, we assess the roles of geometric thickness variations, surface roughness and microstructural effects in determining the EB field and coercivity in nonplanar architectures.

\section{Results}
\subsection{Exchange-biased 3D CoFe/IrMn microwires} 
\begin{figure*}[ht!]

    \includegraphics[width=1.0\linewidth]{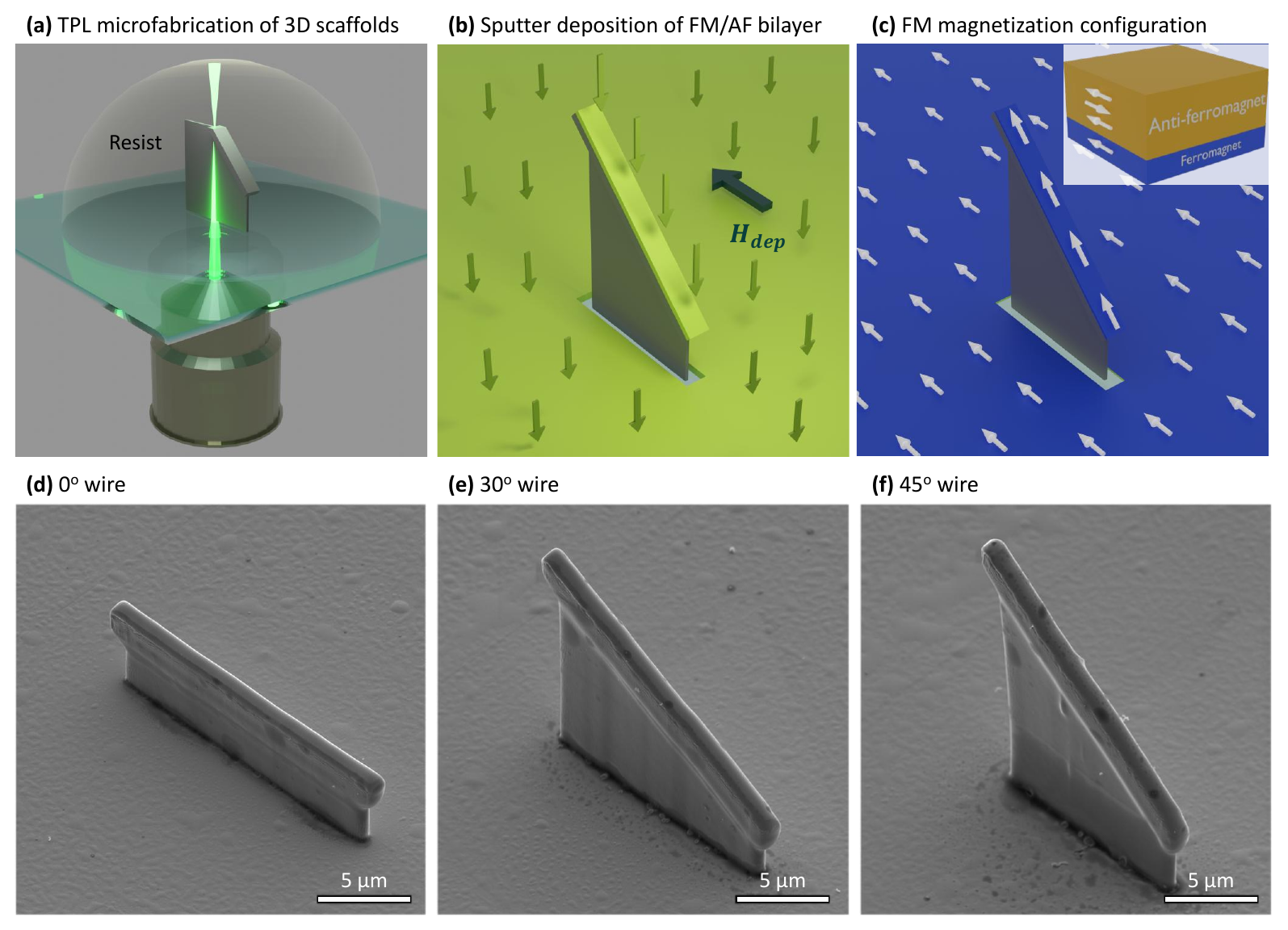}
     \centering

\caption{\textbf{Fabrication of exchange-biased 3D CoFe/IrMn microwires.}
\textbf{(a)} Schematic illustration of the TPL fabrication of 3D microwire scaffolds on a glass substrate, polymerization occurring at the focus point of the laser within the photoresist.
\textbf{(b)} Schematic of the magnetron sputter deposition performed in the presence of an in-plane magnetic field applied along $x$-axis (see Fig.~\ref{fig:main_figure}a for the coordinate systems).
\textbf{(c)} Illustration of the resulting possible CoFe magnetization orientation in planar film and the 3D wire as a result of deposition in the presence of a magnetic field, followed by the IrMn deposition. The inset shows schematically the possible spin orientations of the CoFe/IrMn bilayer.
\textbf{(d-f)} SEM images of the 3D microwires with inclinations of 0°, 30° and 45° (with respect to the substrate) after the deposition of the stack.}
\label{fig:fabrication}
\end{figure*}

The 3D microwire scaffolds are fabricated using TPL (Fig.~\ref{fig:fabrication}a). Each scaffold consists of a flat top segment and a narrower inclined bottom segment, forming a T-shaped cross-section with a width ratio of 1:3 between the bottom and top parts. This T-shape geometry prevents continuity of subsequently deposited sputter films from the top facet to the sidewalls, ensuring that the FM/AF bilayer forms an isolated wire rather than an extended coating over the entire 3D structure. Wire scaffolds are fabricated with inclination angles of $0^\circ$, $30^\circ$ and $45^\circ$ relative to the substrate, a lateral width of 1 $\mu\mathrm{m}$ (Fig.~\ref{fig:main_figure}b) and a constant length of 20 $\mu\mathrm{m}$. The 1 $\mu$m width is chosen to observe any potential patterning effects while remaining well above the resolution limit of TPL, ensuring mechanical stability and sufficient width for atomic force microscopy (AFM) measurements of nanoscale surface roughness (here restricted to the $0^\circ$ wires).

The exchange-biased 3D wires are realized by depositing the stack Ta10/Pt2/CoFe$X$/IrMn$Y$/Pt5 (thicknesses in nm) onto the 3D wire scaffolds via magnetron sputtering (Fig.~\ref{fig:fabrication}b). Here, $X$ and $Y$ denote the variable thicknesses of the CoFe and IrMn layers, respectively. The Ta10/Pt2 bilayer serves as a seed layer to promote high-quality growth of the subsequent layers, while the Pt5 acts as a protective capping layer. The deposited layer stacks Ta10/Pt2/CoFe$X$/IrMn$Y$/Pt5, hereafter referred to as ``CoFe$X$/IrMn$Y$''. Deposition is performed under an external in-plane magnetic field of 50 mT applied along the $x$-axis (see Fig.~\ref{fig:main_figure}a for the coordinate systems). The applied field is expected to align the interfacial FM spins (Fig.~\ref{fig:fabrication}c) and sets the orientation of the uncompensated AF interfacial spins, thereby establishing the unidirectional anisotropy and the EB field.~\cite{Berkowitz2005} This eliminates the need for high-temperature annealing of the 3D scaffolds above the Néel temperature of IrMn. SEM images of the resulting exchange-biased 3D wires are shown in Fig.~\ref{fig:fabrication}d--f. Detailed procedures for TPL fabrication and sputter deposition are provided in the Methods section.

\subsection{Effect of tilted 3D geometry on exchange bias}

\begin{figure*}[ht!]
    \centering
    \includegraphics[width=0.90\linewidth]{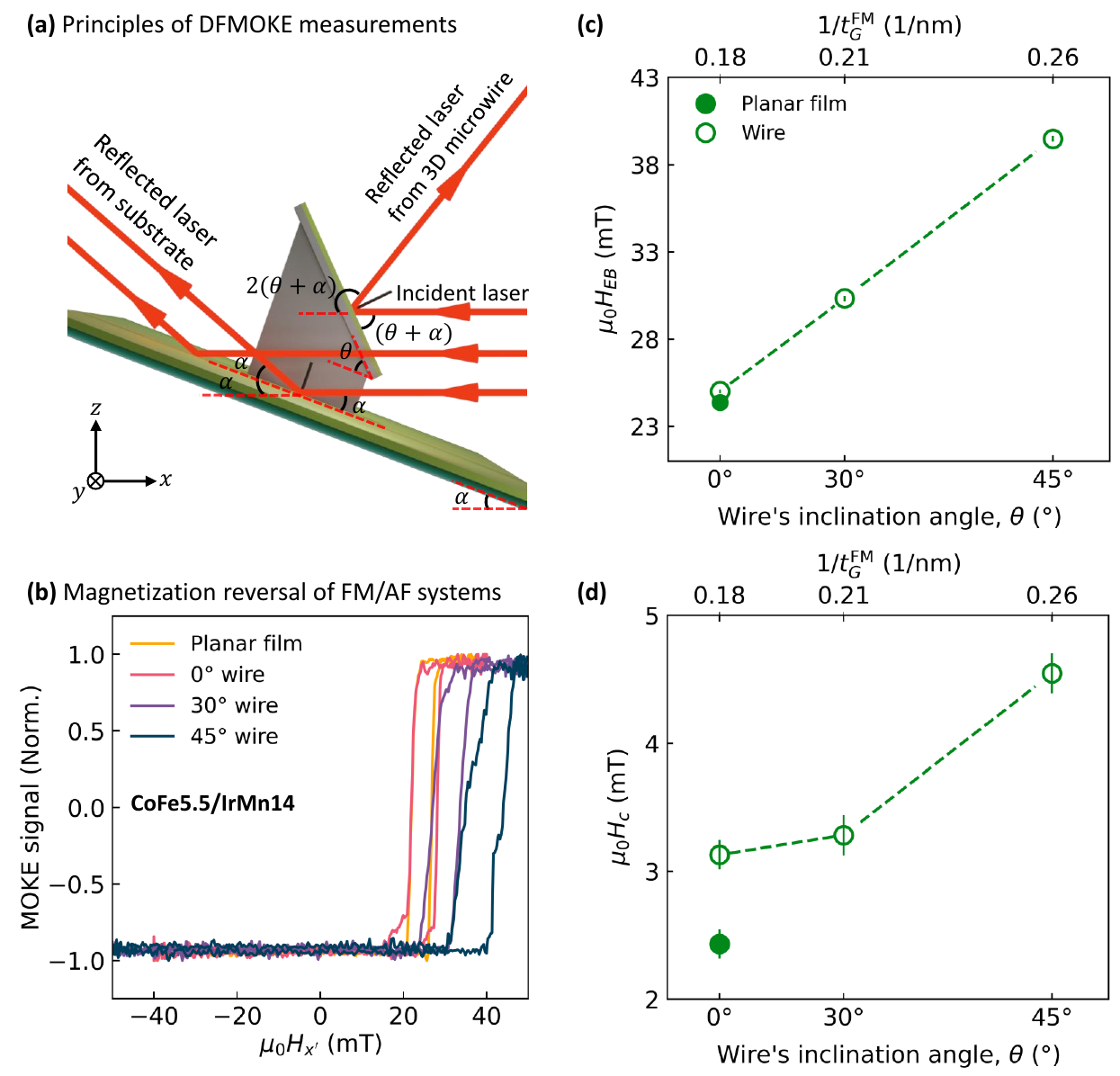}

\caption{\textbf{Probing the simultaneous magnetization reversal of CoFe5.5/IrMn14 planar film and 3D microwires using DF-MOKE.}
\textbf{(a)} Schematic of the DF-MOKE setup. \textbf{(b)} Magnetization reversal loops of planar film and wires of length 20~$\mu$m and width 1~$\mu$m. The magnetic field $H_{x'}$, is applied along $x'$-direction (see Fig.~\ref{fig:main_figure}a for coordinate systems).
\textbf{(c)} Exchange bias and \textbf{(d)} coercivity plotted as a function of the wire's inclination angle, $\theta$. Solid symbols represent the data of the planar film, while open symbols represent the data of the individual wires. The top horizontal axis of both panels corresponds to $1/t_{G}^{\mathrm{FM}}$. In panels (c) and (d), the same symbols are used and dashed lines are used to connect the data points.}
    \label{fig:DF-MOKE}
\end{figure*}

The magnetization reversal of individual 3D wires is investigated using DF-MOKE as a function of the magnetic field $H_{x'}$ applied along the local $x'$ axis, which coincides with the long axis of the wires (see Fig.~\ref{fig:main_figure}a for coordinate systems). The measurement configuration is shown in Fig.~\ref{fig:DF-MOKE}a, where a linearly polarized laser beam (Gaussian spot of $\sim$ 7.5 $\mu$m FWHM) reflects simultaneously from the 3D microwire (dark-field) and the surrounding planar film (bright-field), enabling the independent measurement of their magnetic response. Further details of the DF-MOKE setup are provided in the Methods section.

Representative magnetization reversal for the CoFe5.5/IrMn14 planar film and 3D wires are shown in Fig.~\ref{fig:DF-MOKE}b. The measured loops reveal a systematic increase in the loop shift in the horizontal direction with increasing inclination angle. Moreover, the $45^\circ$ wire exhibits a more complex reversal behavior, with potential origins discussed later in this section. The extracted $H_{\mathrm{EB}}$ and coercive fields ($H_{\mathrm{c}}$) are summarized in Fig.~\ref{fig:DF-MOKE}c,d where the bottom $x$-axis corresponds to the inclination angle $\theta$, while the top $x$-axis corresponds to the calculated inverse of the geometric FM thickness $1/t_G^{\mathrm{FM}}$.

\captionsetup[figure]{labelfont=bf}

For flat geometries (i.e., the planar film and the $0^\circ$ wire), $H_{\mathrm{EB}}$ remains unchanged, indicating that patterning the film into micron-sized wires has no measurable influence on the EB. This observation is consistent with previous reports~\cite{Yu2000, Mao1999}, which show that patterning affects the FM/AF interfacial coupling only when the element dimensions are comparable to or smaller than the intrinsic magnetic length scales of the FM and AF layers, such as the FM domain-wall width and the AF domain size.

For inclined wires ($30^\circ$ and $45^\circ$), $H_{\mathrm{EB}}$ increases systematically with wire inclination (bottom $x$-axis). Consistent with the dependence of EB on FM thickness, this trend reflects an effective reduction of the FM thickness caused by the directional nature of sputter deposition. As a result, when $H_{\mathrm{EB}}$ is plotted against the inverse geometric FM thickness, $1/t_G^{\mathrm{FM}}$ (top $x$-axis), $H_{\mathrm{EB}}$ exhibits a linear dependence, confirming that the reduction of the FM thickness is the primary factor responsible for the observed enhancement. This observation holds irrespective of the fact that geometric AF thickness decreases with inclination and the film microstructure may differ slightly due to oblique-angle deposition~\cite{Ali2021}. The AF layer thus seems to remain well above the thickness required for stable EB~\cite{Nogues1999, Ali2003, Chen2014, Yu2000}, thus making its contribution to the observed change in $H_{\mathrm{EB}}$ negligible. The influence of the reduction in geometric AF thickness on $H_{\mathrm{EB}}$ is seen in other FM/AF systems and is discussed in the next section (Fig.~\ref{fig:summary}b). Microstructural effects are discussed later in this section.

We focus our attention now on the dependence of the coercive field ($H_{\mathrm{c}}$) on the wire's inclination angle (Fig.~\ref{fig:DF-MOKE}d). A small increase in $H_{\mathrm{c}}$ is observed for the $0^\circ$ wire relative to the planar film. Furthermore, a steady increase of $H_c$ with $\theta$ is observed. As reported by Yu \emph{et al.}~\cite{Yu2000}, coercivity in exchange-biased elements is primarily governed by FM/AF interfacial coupling, with magnetization reversal initiated via random domain nucleation within the interior of the elements, accompanied by strongly pinned domain walls, in contrast to unbiased patterned structures, where nucleation typically initiates at the edges due to demagnetizing fields. Thus, the nucleation and pinning are typically dominated by spatial fluctuations in the FM--AF exchange coupling at the interface, occurring on length scales smaller than the micron-sized elements. Consequently, both $H_{\mathrm{c}}$ and $H_{\mathrm{EB}}$ are found to be largely independent of the lateral dimensions and to approximately scale with $1/t^{\mathrm{FM}}$, revealing their interfacial origin. Size effects would become relevant only at deep submicron scales approaching the characteristic AF domain size. Consistent with this model, the slight increase in $H_{\mathrm{c}}$ for the $0^\circ$ wire can be attributed to a local reduction in the FM thickness at the wire edges, arising from the combination of rounded edges (as observed in the SEM image, Fig.~\ref{fig:fabrication}d) and the directional nature of the sputter deposition. This local thinning of the FM layer likely modifies the FM--AF interfacial coupling and enhances domain-wall pinning.

The coercivity of the inclined wires increases with inclination angle, consistent with the reduction in the geometric FM thickness. This behavior aligns with previous studies, which report an increase in coercivity of a few mT per nanometer decrease in FM thickness~\cite{Yu2000, Nogues1999}. Since the geometric AF thickness on the $45^\circ$ wire remains $\approx$ 9.9 nm, the influence of reduced AF thickness on the coercivity is expected to be weak; however, any residual contribution would tend to enhance coercivity~\cite{Jungblut1994_OrientationalEB, Chen2014, Yu2000}. The influence of the reduction in geometric AF thickness in $H_{\mathrm{c}}$ is presented and discussed in the next section (Fig.~\ref{fig:summary}c).

Oblique-angle deposition is known to modify the film microstructure, promoting columnar growth and increased interface roughness~\cite{Ali2021}. Fleischmann \emph{et al.}~\cite{Fleischmann2010_CoO_Fe_Roughness} demonstrated that enhanced interface roughness in exchange-biased FM/AF systems increases coercivity by introducing additional pinning sites, while the $H_{EB}$ exhibits a nonmonotonic dependence on roughness due to competing effects of uncompensated spins and interfacial disorder. As discussed in the following section, in our case, we understand the increase in $H_{\mathrm{EB}}$ for inclined wires to be primarily attributed to a reduction in FM thickness, indicating that microstructural changes have a minor effect on $H_{\mathrm{EB}}$. Nevertheless, the more complex magnetization reversal observed for the 45° wire suggests enhanced domain-wall pinning, which may arise from either the modified film microstructure or the reduced FM thickness locally at the wire edges.

\subsection{Role of surface roughness and tilted geometry on bilayer growth and magnetic performance}  

As discussed previously, $H_{\mathrm{EB}}$ for the CoFe5.5/IrMn14 wires increases linearly with $1/t_{G}^{\mathrm{FM}}$. To verify the reliability of the calculated geometric FM thicknesses ($t_{G}^{\mathrm{FM}}$), two additional FM/AF bilayers with different nominal FM thicknesses are fabricated under similar deposition conditions: CoFe7/IrMn14 deposited on a substrate with 3D wire scaffolds, and CoFe8.5/IrMn14 deposited on a planar substrate.  

To investigate the influence of AF thickness, two extra FM/AF bilayers with different nominal AF thicknesses are prepared: CoFe5.5/IrMn5 deposited on a substrate with 3D wire scaffolds and CoFe5.5/IrMn3 deposited on a planar substrate. Together, these five FM/AF systems form FM series and AF series, with the CoFe5.5/IrMn14 stack common to both series.

\begin{figure*}[h!]
    \centering
    \includegraphics[width=0.90\linewidth]{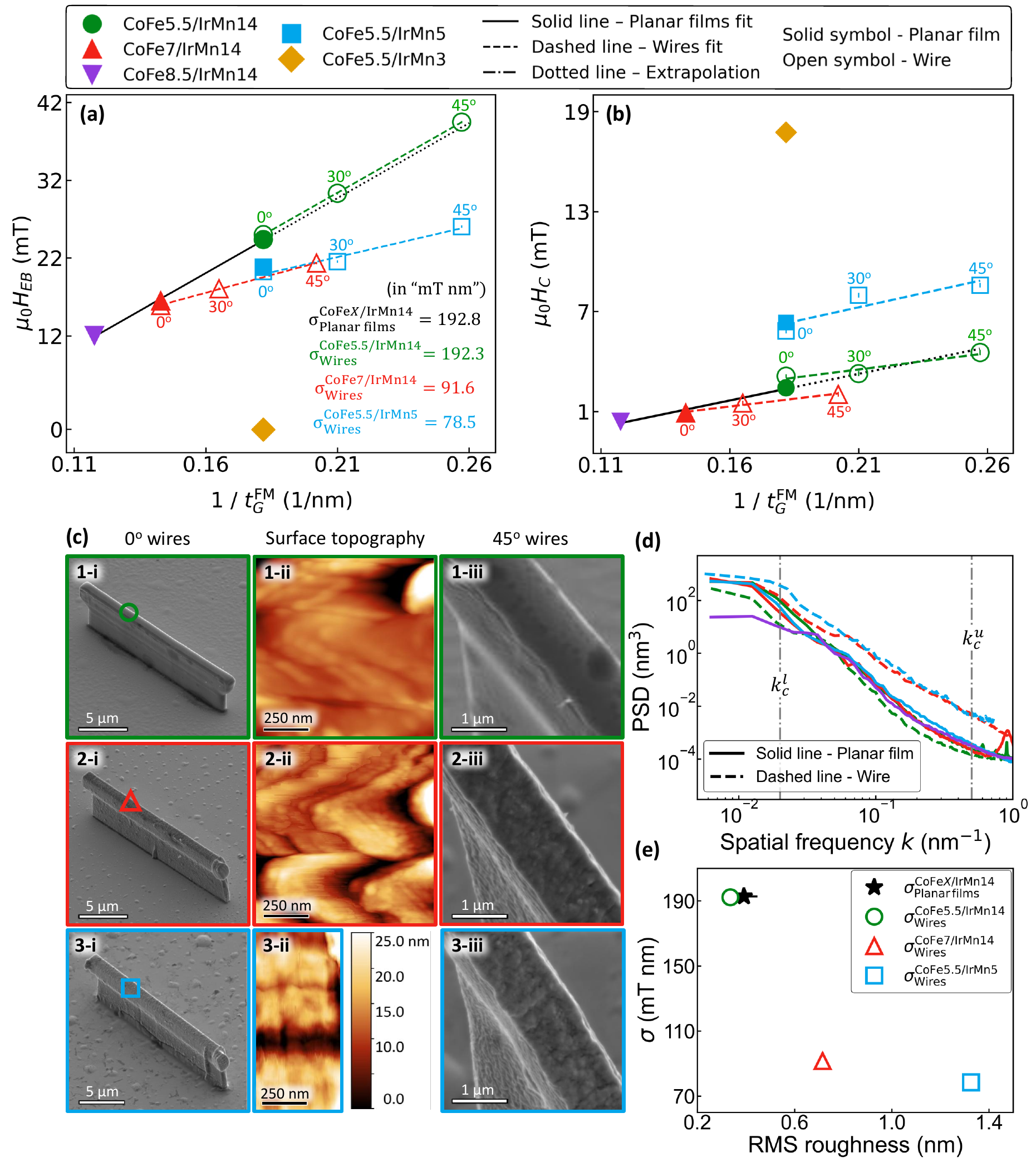}
    \caption{\textbf{Exchange bias and coercivity as a function of wire geometry and surface roughness.} 
\textbf{(a)} EB and \textbf{(b)} Coercivity as a function of inverse geometric FM thickness ($1/t_{G}^{\mathrm{FM}}$) for FM- and AF-series. Solid symbols correspond to planar film data, while open symbols represent wire data. Dashed lines indicate linear fits to the wire data at the three inclination angles, whereas the solid black line shows the linear fit to the planar film data in the FM series, with the dotted black line representing its extrapolation to lower $t_{0}^{\mathrm{FM}}$ values. $\sigma_{\mathrm{Planar\ films/Wires}}^{\mathrm{CoFe}\mathit{X}/\mathrm{IrMn}\mathit{Y}}$ (in $\mathrm{mT\,nm}$) denotes the slope obtained from the linear fit to the ``Planar films/Wires'' data points of the $\mathrm{CoFe}\mathit{X}/\mathrm{IrMn}\mathit{Y}$ bilayer. The labels $0^\circ$, $30^\circ$ and $45^\circ$ associated with the wire data points indicate the corresponding wire inclination angles. \textbf{(c)} SEM images of 0° wires (first column) for \textbf{(1)} CoFe5.5/IrMn14, \textbf{(2)} CoFe7/IrMn14 and \textbf{(3)} CoFe5.5/IrMn5, with corresponding AFM topographies (second column) acquired at indicated marker positions. The height color scale is shared across AFM images. The third column shows magnified views of 45° wires. 
\textbf{(d)} Power spectral density (PSD) vs. spatial frequency $k$, with planar films shown as solid lines and wires as dashed lines. \textbf{(e)} Slopes, extracted from linear fits of $H_{\mathrm{EB}}$ versus $1/t_{G}^{\mathrm{FM}}$, are plotted as a function of RMS roughness. Roughness data correspond to the planar geometries: the $0^\circ$ wire and the average of the $\mathrm{CoFe}\mathit{X}/\mathrm{IrMn}14$ planar films, with the standard deviation from this averaging used as the error bar. Marker symbols and colors, as indicated in the legend at the top, are consistent across all panels.}
    \label{fig:summary}   
\end{figure*}

The magnetization reversal loops of the four additionally prepared FM/AF stacks are shown in Supporting Information Fig. S1. The extracted $H_{\mathrm{EB}}$ and $H_{\mathrm{c}}$ (together with the one for CoFe5.5/IrMn14 presented in Fig.~\ref{fig:DF-MOKE}) are summarized in Fig.~\ref{fig:summary}. Each FM/AF bilayer grown on a substrate with wire scaffolds contributes to four data points: one from the planar film and three from wires with inclination angles of $0^\circ$, $30^\circ$ and $45^\circ$. In contrast, FM/AF bilayers deposited only on planar substrates contribute only to a single data point. In Fig.~\ref{fig:summary}, solid symbols denote planar film data and open symbols correspond to wire data. Dashed lines represent linear fits to the wire data ($H_{\mathrm{EB}}$ vs. $1/t_{G}^{\mathrm{FM}}$), the solid black line represents the linear fit for the planar films ($H_{\mathrm{EB}}$ vs. $1/t_{0}^{\mathrm{FM}}$) and the dotted black line indicates the extrapolation of the planar film fit to lower $t_{0}^{\mathrm{FM}}$.

\subsubsection{Exchange bias} Fig.~\ref{fig:summary}a summarizes the EB fields for all FM/AF
systems. Focusing first on the flat geometries of the FM series (CoFe$X$/IrMn14 bilayers), i.e., the planar film and the $0^\circ$ wire, CoFe7/IrMn14 systems exhibit similar $H_{\mathrm{EB}}$, confirming that patterning has no influence on the EB. Furthermore, $H_{\mathrm{EB}}$ decreases as the nominal CoFe thickness $t_{0}^{\mathrm{CoFe}}$ is increased from 5.5 to 7 and 8.5 nm, consistent with the expected inverse scaling with FM thickness. 

The slope obtained from the linear fit to the planar film data (black line), $\sigma_{\mathrm{Planar\,films}}^{\mathrm{CoFe\mathit{X}/IrMn14}} = 192.8~\mathrm{mT\,nm}$, is in good agreement with that extracted from the 3D CoFe5.5/IrMn14 wires (green, dashed line), $\sigma_{\mathrm{Wires}}^{\mathrm{CoFe5.5/IrMn14}} = 192.3~\mathrm{mT\,nm}$. Moreover, the data points of this wire series lie in direct continuation of the planar film trend, without any significant vertical offset (comparison between the green dashed line and the black dotted line). indicating that $H_{\mathrm{EB}}(\text{Planar\,films}) = H_{\mathrm{EB}}(\text{Wires})$, provided that the FM thickness is identical and the AF layer thickness exceeds the critical value required to establish a stable EB.
The close agreement between the slopes, which represent the averaged interfacial exchange coupling strength, confirms the reliability of the calculated geometric FM thicknesses $t_{G}^{\mathrm{FM}}$ and demonstrates that the FM/AF interface quality in bilayers deposited on 3D wire scaffolds is comparable to that of planar films. Furthermore, the identical slopes for the CoFe5.5/IrMn14 wires and the planar films substantiate the assumption made in the previous section that any reduction in geometric AF thickness does not significantly alter $H_{\mathrm{EB}}$ for the CoFe5.5/IrMn14 wires inclined at $30^\circ$ and $45^\circ$.
Finally, the extracted slope ($\sigma \approx 192.5~\mathrm{mT\,nm}$) can be regarded as a characteristic physical parameter describing both the interfacial quality of our FM/AF bilayer and the condition that the AF thickness is sufficiently large to support a stable EB.

$H_{\mathrm{EB}}$ of the 3D CoFe7/IrMn14 wires also exhibits a linear dependence with $1/t_{G}^{\mathrm{FM}}$, but with a noticeably smaller slope, $\sigma_{\mathrm{wires}}^{\mathrm{CoFe\,7/IrMn\,14}} = 91.6~\mathrm{mT\,nm}$ (see red dashed line). Since $H_{\mathrm{EB}}$ for the $0^\circ$ wire (open red triangle) is nearly identical to that of the planar film (solid red triangle), the reduced slope is due to a lower $H_{\mathrm{EB}}$ observed for the $30^\circ$ and $45^\circ$ with and the possible reason discussed later in this section.

Focusing now on the flat geometries of the AF series (CoFe5.5/IrMn$Y$ bilayers, see blue and yellow symbols, either solid or open ones marked with $0^\circ$), $H_{\mathrm{EB}}$ decreases for both the CoFe5.5/IrMn5 planar film and the corresponding $0^{\circ}$ wire as the nominal AF thickness $t_{0}^{\mathrm{AF}}$ is reduced from 14 to 5 nm. Again, the contribution of patterning to the EB seems negligible. Moreover, upon further reducing $t_{0}^{\mathrm{AF}}$ to 3 nm (yellow solid symbol), $H_{\mathrm{EB}}$ vanishes for the CoFe5.5/IrMn3 planar film.

These observations are consistent with previous reports~\cite{Xi2000, Ali2003, Nogues1999}, which indicate that the EB is governed by irreversible (pinned) AF interfacial moments. For AF thicknesses below a critical value $t_{\mathrm{cr}}^{\mathrm{AF}}$, the AF layer lacks sufficient anisotropy and domain stability to sustain a pinned interfacial spin configuration, resulting in a vanishing $H_{\mathrm{EB}}$. As the AF thickness increases beyond $t_{\mathrm{cr}}^{\mathrm{AF}}$, stable AF domains can form at the interface, giving rise to a rapidly increasing $H_{\mathrm{EB}}$, which subsequently approaches a saturated or weakly non-monotonic regime at larger thicknesses.

The slope of the linear fit performed on the CoFe5.5/IrMn5 wires, $\sigma_{\mathrm{wires}}^{\mathrm{CoFe5.5/IrMn5}} = 78.5~\mathrm{mT\,nm}$ (see blue dashed line)), is further reduced compared to that obtained for the CoFe7/IrMn14 wires (red dashed line). This decrease of $H_{\mathrm{EB}}$ in CoFe5.5/IrMn5 tilted wires, more pronounced than that observed in CoFe7/IrMn14 tilted wires, indicates two possible contributing mechanisms. First, there is a contribution from the same mechanism that reduces $H_{\mathrm{EB}}$ in CoFe7/IrMn14 wires (discussed later in this section). Second, the additional reduction is attributed to the effective decrease in AF thickness in inclined wires, as evidenced by the reduction of $H_{\mathrm{EB}}$ when the AF thickness is lowered from 14 to 5 nm and further to 3 nm. With only 5 nm in the CoFe5.5/IrMn5 planar film, this AF layer thickness is expected to be rather close to the critical AF thickness and decreasing it further for a larger inclination of wires will continuously decrease $H_{EB}$.

In our study of EB, a finite $H_{\mathrm{EB}}$ is observed in the low AF thickness regime at a geometric AF thickness $t_{G}^{\mathrm{AF}}$ $\approx$ 3.5 nm (CoFe5.5/IrMn5, $45^{\circ}$ wire), whereas the EB vanishes for $t_{0}^{\mathrm{AF}}$ $\approx$ 3 nm (CoFe5.5/IrMn5 planar film). This behavior indicates that the critical AF thickness, $t_{\mathrm{cr}}^{\mathrm{AF}}$, lies between 3 and 3.5 nm.
In the medium AF thickness regime, $H_{\mathrm{EB}}$ is sensitive to the AF thickness at 5 nm (CoFe5.5/IrMn5, $0^{\circ}$ wire), whereas it becomes thickness-independent for thicknesses of $\approx$ 9.9 nm (CoFe5.5/IrMn14, $45^{\circ}$ wire). This suggests that the AF thickness required to saturate the $H_{\mathrm{EB}}$ lies in between 5 and 10 nm.

\subsubsection{Coercivity} Fig.~\ref{fig:summary}b summarizes the coercive fields for all FM/AF systems. 
The coercivity behavior of the CoFe5.5/IrMn14 system is discussed in the previous section, Fig.~\ref{fig:DF-MOKE}d.
For the CoFe7/IrMn14 and CoFe5.5/IrMn5 systems, comparison between the planar films and the corresponding $0^{\circ}$ wires shows no significant difference in $H_{\mathrm{c}}$, confirming that lateral patterning into micron-sized wires has a negligible effect on coercivity.

For inclined wires ($30^{\circ}$ and $45^{\circ}$), $H_{\mathrm{c}}$ increases monotonically with $1/t_{G}^{\mathrm{FM}}$, indicating that the coercivity is primarily governed by the reduction of the geometric FM thickness. 
However, the increase in coercivity is notably stronger for the CoFe5.5/IrMn5 wires, indicating the presence of an additional contribution. This contribution likely originates from the reduced AF thickness and is negligible in CoFe5.5/IrMn14 and CoFe7/IrMn14 systems, where the AF layer is sufficiently thick.

In planar films, $H_{\mathrm{c}}$ increases systematically with both increasing $1/t_{0}^{\mathrm{FM}}$ (nominal FM thickness reduced from 8.5 to 7 to 5.5 nm) and increasing $1/t_{0}^{\mathrm{AF}}$ (nominal AF thickness reduced from 14 to 5 to 3 nm). The increase in coercivity with decreasing FM thickness is consistent with previous reports~\cite{Nogues1999, Yu2000} and its origin has been discussed in the previous section. Furthermore, the increase in $H_{\mathrm{c}}$ with decreasing AF thickness is also consistent with earlier studies~\cite{Xi2000, Ali2003, Nogues1999}. According to these studies, the coercivity of the FM layer first increases when it is interfaced with a thin AF layer compared to when it is isolated. This enhancement typically reaches a maximum near the critical AF thickness, $t_{\mathrm{cr}}^{\mathrm{AF}}$ and arises from the coupling of the FM layer to reversible or rotatable AF interfacial spins, which act as effective pinning centers and hinder domain-wall motion. Upon further increasing the AF layer thickness, the interfacial AF moments become more and more pinned and less rotatable. Consequently, $H_{\mathrm{c}}$ generally decreases with increasing AF thickness above $t_{\mathrm{cr}}^{\mathrm{AF}}$ and stabilizes at a moderately elevated value. 

In agreement with this behavior, our measurements show that at $t_{0}^{\mathrm{AF}}$ = 3 nm, the coercivity reaches a maximum (among all studied AF thicknesses), indicating that the critical AF thickness, $t_{\mathrm{cr}}^{\mathrm{AF}}$, lies near 3 nm (as evidenced from the $H_{\mathrm{EB}}$ measurements as well). For AF layers thinner than this threshold, the coercivity is expected to decrease.

\subsubsection{Surface morphology} Fig.~\ref{fig:summary}c presents SEM and AFM images of 3D wires for three FM/AF systems: CoFe5.5/IrMn14 wires (scaffold fabricated with resist 1), CoFe7/IrMn14 and CoFe5.5/IrMn5 wires (scaffold fabricated with resist 2). The SEM images of the $0^\circ$ wires show that scaffold quality decreases when using resist 2. AFM imaging of the top surfaces of the $0^\circ$ wires (positions indicated in the first column with an open symbol) confirms an increased surface roughness for CoFe7/IrMn14 and CoFe5.5/IrMn5 $0^\circ$ wires compared to the CoFe5.5/IrMn14 $0^\circ$ wire. Although AFM could not be performed on the $30^\circ$ and $45^\circ$ wires due to their inclination, SEM images (third column; magnified views, with full images shown in Fig.~\ref{fig:fabrication}e,f and in Supporting Information, Fig. S2b-iii,iv and S2c-iii,iv) reveal a similar trend in surface roughness as observed for the $0^\circ$ wires. Moreover, the SEM images qualitatively indicate that the wires at $0^\circ$, $30^\circ$ and $45^\circ$ maintain comparable surface features.

Fig.~\ref{fig:summary}d shows the power spectral density (PSD) functions, which provide a quantitative description of the spatial frequency components of surface roughness, for FM/AF systems exhibiting EB (i.e., excluding the CoFe5.5/IrMn3 planar film). The PSD functions were derived from height profiles obtained from AFM images, measured on the planar films (solid lines; corresponding images are shown in Supporting Information, Fig. S2a-ii--c-ii) and on the top surfaces of the $0^\circ$ wires (dashed lines; images shown in Fig.~\ref{fig:summary}c1-ii--c3-ii). RMS roughness values were then calculated from these PSD plots in order to extract the contribution of roughness associated with specific spatial frequency ranges. On the PSD functions, the lower and upper limits of the cutoff wave vector $k_{\mathrm{c}}$, denoted as $k^{l}_{c}$ and $k^{u}_{c}$ (indicated by vertical dash-dotted lines), correspond to wavelengths of 50 nm and 2 nm, respectively. The upper limit of the wavevector is defined by the AFM spatial resolution, whereas the lower limit corresponds to the magnetic characteristic length of the FM layer, which is taken as the in-plane domain-wall width of $\approx$ 50 nm for a 5.5 nm CoFe film. \cite{Hubert1998} Roughness corresponding to a wavevector lower than the lower limit represents long-wavelength modulations, which may not significantly affect the EB, as it is based on a short-range interaction.

\subsubsection{Correlation of exchange bias with surface roughness of 3D geometry} In Fig.~\ref{fig:summary}e, $\sigma$ (extracted from the linear fits in Fig.~\ref{fig:summary}a, representing the averaged interfacial exchange energy density) is plotted as a function of the RMS roughness for planar geometries. For the planar films, $\sigma_{\mathrm{Planar\,films}}^{\mathrm{CoFe}\mathit{X}/\mathrm{IrMn}14}$ is plotted against the roughness averaged over the three $\mathrm{CoFe}\mathit{X}/\mathrm{IrMn}14$ films, as their PSD profiles within the selected wave vector range are comparable. For the wire geometries, the slopes $\sigma_{\mathrm{Wires}}^{\mathrm{CoFe}\mathit{X}/\mathrm{IrMn}\mathit{Y}}$ are plotted against the roughness of the corresponding $0^\circ$ wires.
The plot shows that for low roughness (e.g., the CoFe5.5/IrMn14 $0^\circ$ wire), $\sigma_{\mathrm{Wires}}^{\mathrm{CoFe5.5/IrMn14}}$ closely matches $\sigma_{\mathrm{Planar\,films}}^{\mathrm{CoFe\mathit{X}/IrMn14}}$, confirming the high quality of the TPL-printed (resist 1) wire scaffolds for magnetic multilayer integration. As the roughness increases (e.g., in CoFe7/IrMn14 and CoFe5.5/IrMn5 $0^\circ$ wires), although the EB of the $0^\circ$ wires remains largely unaffected (matching that of the corresponding planar films), $\sigma_{\mathrm{Wires}}^{\mathrm{CoFe7/IrMn14}}$ and $\sigma_{\mathrm{Wires}}^{\mathrm{CoFe5.5/IrMn5}}$ are reduced. In the case of CoFe5.5/IrMn5 wires, the reduction in $\sigma$ is further influenced by the decreased AF layer thickness.

Thus, we understand that the reduced $\sigma_{\mathrm{Wires}}^{\mathrm{CoFe7/IrMn14}}$ and $\sigma_{\mathrm{Wires}}^{\mathrm{CoFe5.5/IrMn5}}$, associated with the decreased $H_{\mathrm{EB}}$ observed in the rougher CoFe7/IrMn14 and CoFe5.5/IrMn5 $30^\circ$ and $45^\circ$ wires (Fig.~\ref{fig:summary}a,b), originates primarily from directional sputter deposition onto inclined, rough surfaces. In particular, long-wavelength roughness can induce shadowing effects during deposition, leading to local thickness inhomogeneities at the wire surface. These inhomogeneities weaken the interfacial exchange coupling and consequently reduce $H_{\mathrm{EB}}$.

\subsection{Conclusions} This work demonstrates the realization of exchange-biased 3D CoFe/IrMn microwires by integrating TPL-fabricated microwire scaffolds of inclination angles up to $45^{\circ}$ with magnetron sputtered CoFe/IrMn bilayer. DF-MOKE measurements reveal a systematic increase of $H_{\mathrm{EB}}$ and $H_{\mathrm{c}}$ with wire inclination, which is primarily attributed to the geometry-induced reduction of the FM layer thickness, a result of the directional deposition on a 3D tilted geometry. Analysis of surface morphology and the variation of $H_{\mathrm{EB}}$ with both geometric FM thicknesses ($t^{\mathrm{FM}}_{G}$) and nominal FM thicknesses ($t^{\mathrm{FM}}_{0}$) indicates that the interface quality of FM/AF bilayer deposited on smooth 3D wire scaffolds remains largely unaffected by the 3D geometry or the film microstructure, highlighting the compatibility of 3D geometries with multilayered stacks with functional interfaces grown by physical vapor deposition methods.\\
While $H_{\mathrm{EB}}$ remains largely insensitive to micron-scale lateral patterning and to RMS roughnesses up to $\approx$1.5 nm in $0^{\circ}$ wires, a reduction in $H_{\mathrm{EB}}$ is observed for the rougher, inclined wires. This decrease is attributed to the combined effects of the 3D scaffold geometry, surface roughness and directional deposition, which likely give rise to thickness inhomogeneities in the deposited layers due to shadowing. 
Additionally, the coercivity is also understood to be primarily governed by thickness variations in the FM and AF layers, rather than by shape anisotropy or glancing-angle-induced modifications of the film microstructure.

In summary, this work systematically examines how the thickness of PVD-deposited CoFe/IrMn bilayers varies with microwire inclination and how the combined effects of surface roughness, 3D geometry and directional deposition influence the magnitude of the EB. These findings are crucial for preparing complex 3D geometries with magnetic multilayers, where variations in scaffold surface quality and inclination can significantly impact interfacial magnetic properties.
Future studies could investigate the role of 3D geometry on other interfacial magnetic phenomena, such as perpendicular magnetic anisotropy, RKKY coupling and Dzyaloshinskii–Moriya interactions. Finally, extending such studies to nanoscale 3D scaffolds fabricated via other methods with better spatial resolution and surface roughness, such as focused electron beam-induced deposition (FEBID), could provide fundamental insights into the influence of nanoscale 3D geometrical tuning on interfacial magnetic properties.

\section{Methods}

\subsection{Fabrication} \textbf{Two-Photon Lithography (TPL):} TPL is a direct laser-writing technique based on the nonlinear optical process of two-photon absorption. \cite{Askey2025} A tightly focused picosecond laser induces polymerization exclusively within the focal volume, enabling the fabrication of 3D microscale geometries with sub-micrometer resolution. In this work, TPL is used to fabricate \(20~\mu\text{m}\)-long wires of width  \(1~\mu\text{m}\), fabricated at inclination angles of \(0^\circ\), \(30^\circ\) and \(45^\circ\). The structures are fabricated using a Microlight3D system equipped with a 532-nm pulsed laser (base power 16 mW) and a 100X oil-immersion objective with a numerical aperture (NA) of 1.25. 3D models are designed in Blender, exported as STL files and imported into the Microlight3D control software, where they are sliced along the $y$-direction (see Fig.~\ref{fig:main_figure}a) and the highest-quality slicing is chosen. This workflow yields a voxel size of approximately \(200~\text{nm}\) in the lateral ($x$-$y$) plane and \(500~\text{nm}\) along the $z$-direction.
Two proprietary negative-tone photoresists are used for microfabrication: Green-A (referred to as “resist 1” in the Results section), employed for scaffolds subsequently coated with CoFe5.5/IrMn14 and OrmoBio (referred to as “resist 2” in the Results section), employed for scaffolds subsequently coated with CoFe7/IrMn14 and CoFe5.5/IrMn5. Fabrication is performed in transmission mode by using a drop of photoresist on a cleaned \(150~\mu\text{m}\)-thick borosilicate glass substrate, focusing the laser at the glass-resist interface using the system’s camera and scanning the substrate with a three-axis piezo stage. A laser power of 8\% and a scanning speed of $20~\mu\mathrm{m/s}$ are used for the exposure of both resists. However, these printing parameters (laser power and scan speed) are optimized for the Green-A resist. After printing, the samples are developed in OrmoDev for 20 minutes and subsequently rinsed in isopropanol for 5 minutes. As the same fabrication parameters are applied to OrmoBio without optimization for this particular resist, the structures printed with proprietary OrmoBio exhibit increased surface roughness.\\ 
\\

\textbf{Magnetron Sputtering:} The exchange-biased system Ta10/Pt2/CoFeX/IrMnY/Pt5 (thicknesses in nm) is deposited in a single run using an AJA International PVD system. The substrate holder remained stationary during the deposition process. The base pressure prior to deposition was \(2.1 \times 10^{-7}~\text{mbar}\), and the Ar sputtering pressure during deposition was maintained at \(1 \times 10^{-3}~\text{mbar}\). The deposition rates for Ta, Pt, CoFe and IrMn are \(1.34\), \(0.81\), \(0.75\) and \(3.56~\text{\AA s}^{-1}\), respectively. The applied sputtering powers for the targets Ta, Pt, CoFe and IrMn are 102, 20, 20 and 102 W, respectively. The FM and AF layer thicknesses are varied between run-to-run depositions (keeping all other parameters identical), depositing a total of five exchange-biased systems: CoFe5.5/IrMn14, CoFe7/IrMn14, CoFe8.5/IrMn14, CoFe5.5/IrMn5 and CoFe5.5/IrMn3.
 All depositions are performed in the presence of an in-plane magnetic field of $50~\mathrm{mT}$ applied along the wire axis ($-x$-axis). On inclined 3D wires, the effective magnetic field along the local wire axis ($x'$-axis) corresponds to the projection of this field onto the inclined surface. It therefore decreases with increasing inclination angle, reaching $35.4~\mathrm{mT}$ for wires inclined at $45^\circ$. Owing to the soft magnetic nature of CoFe, this reduced field is still sufficient to saturate the FM layer~\cite{Chai2012}. This assumption is further supported by the observed EB behavior of the CoFe5.5/IrMn14 systems on 3D wires: the slope of the linear fit to $H_{\mathrm{EB}}$ as a function of $1/t_{G}^{\mathrm{FM}}$ is in good agreement with that obtained from the linear fit of $H_{\mathrm{EB}}$ versus $1/t_{0}^{\mathrm{FM}}$ for planar films. This agreement confirms that the effective magnetic field during deposition is sufficient to establish a well-defined EB even on inclined 3D geometries.

\subsection{Surface roughness characterization} 
\textbf{Scanning electron microscopy (SEM):} SEM is performed using a TESCAN Clara SEM operated in conventional high-vacuum mode. To optimize surface morphology contrast, the acceleration voltage is set to \(3~\text{kV}\). A beam current of \(100~\text{nA}\) is used to maximize the signal-to-noise ratio while minimizing hydrocarbon-induced surface contamination. All images are acquired using a standard Everhart-Thornley secondary electron detector (ETD) with a working distance of approximately \(10~\text{mm}\).\\
\\
\textbf{Atomic force microscopy (AFM):} Atomic force microscopy (AFM) measurements are performed using an Asylum Research Cypher AFM equipped with commercial PPP-NCH probes (Nanosensors). The probes feature a tip radius of \(<10~\text{nm}\), a stiffness of \(42~\text{N/m}\) and a resonance frequency of \(330~\text{kHz}\). Topographies are recorded in standard tapping mode.

\subsection{Magnetic characterization}
\textbf{Dark-field magneto-optical Kerr effect (DF-MOKE) magnetometry:} Magneto-optical properties are measured using a home-built DF-MOKE system equipped with a p-polarized red laser diode ($\lambda = 658~\text{nm}$). The system utilizes a hexapole electromagnet capable of applying a magnetic field in an arbitrary direction up to $60~\text{mT}$. Samples are mounted on a piezo stage capable of translation in the $x$, $y$ and $z$ directions, as well as $360^\circ$ rotation around the $y$-axis. The setup consists of two detection arms, one bright-field and the other dark-field, each equipped with a quarter-wave plate, a half-wave plate, a Wollaston prism and a pair of each photodiodes for differential signal detection. The stage angle set during the measurement, $\alpha$, is $20^\circ$. Light reflected at different angles from the substrate and the individual 3D wire passes through the bright-field and dark-field arms, respectively. The DF-MOKE system is controlled using custom Python-based software. Details of the setup and its components are provided in Ref.~\cite{SanzHernandez2023JAP}.

\subsection{Statistical analysis} The hysteresis loops presented in Figs.~\ref{fig:DF-MOKE} and S1 (in Supporting Information) are the result of averaging 120 magnetization reversal loops. The error bars in the $H_{c}$ and $H_{EB}$ in Figs.~\ref{fig:DF-MOKE}, \ref{fig:summary}a,b, are estimated directly from the numerical determination of the zero-crossing points in the hysteresis loops. Two contributions are considered. First, the finite field resolution of the measurement introduces an intrinsic uncertainty of $\Delta H / 2$, where $\Delta H$ is the mean field step between consecutive data points. Second, noise in the MOKE signal near the reversal point ($|M| < 0.1$) produces an additional uncertainty in the zero-crossing obtained from the linear interpolation. The standard deviation of this noise, $\sigma_M$, is converted into an equivalent field uncertainty by $\sigma_M \, \Delta H$. The total uncertainty in the field values is then obtained by adding both contributions in quadrature:
\[
\sigma_H = \sqrt{\left( \frac{\Delta H}{2} \right)^2 + \left( \sigma_M \, \Delta H \right)^2 }.
\]
This uncertainty is propagated equally to the estimates of $H_{c}$ and $H_{EB}$, as both rely on the determination of the two zero-crossing fields.\\
Small uncertainties in the field magnitudes may have resulted from the field orientation relative to the wire's long axis during both deposition and MOKE measurements; these errors are not included in the error bars.

\section*{Data Availability} The experimental data supporting the findings of this study are available on the TU Wien repository.

\section*{Acknowledgments} We want to thank C. Krien (Leibniz IFW Dresden) for the deposition of magnetic thin films. This work is supported by the European Community under the Horizon 2020 Program, Contract No. 101001290 (3DNANOMAG).\\

\section*{Author Contribution} Conceptualization and Methodology: A.F.P., B.S.; Investigation: B.S., A.S., J.J., S.M.; Resources: A.F.P., V.N., J.J., A.S., S.M., B.S.; Supervision: A.F.P.; Writing - original draft: B.S.; Writing - review and editing: A.F.P., B.S., V.N., J.J., A.S.\\

\section*{Competing Interests} The authors declare that they have no competing interests.

\section*{References}

\bibliographystyle{apsrev4-2}  
\bibliography{referencesnew}      

\clearpage
\onecolumngrid
\appendix

{\centering \Large \textbf{Supporting Information} \par}

\vspace{1em}

{\centering \large \textbf{Geometric dependence of exchange bias in tilted three-dimensional CoFe/IrMn microwires} \par}

\vspace{1em}

{\centering \small {
Balram Singh$^{1}$, Aman Singh$^{2,3}$, Stefan Mikulik$^{1}$, Jakub Jurczyk$^{1}$, Volker Neu$^{2}$, and Amalio Fernández-Pacheco$^{1}$
} \par}

\vspace{0.5em} 

{\centering \small \textit{
$^{1}$ Institute of Applied Physics, TU Wien, 1040 Vienna, Austria\\
$^{2}$ Leibniz Institute for Solid State and Materials Research Dresden, 01069 Dresden, Germany\\
$^{3}$ Institute of Applied Physics, TU Dresden, 01187 Dresden, Germany
} \par}

\vspace{5em}

\setcounter{figure}{0}
\renewcommand{\thefigure}{S\arabic{figure}}

\begin{figure}[h]
    \centering
    \includegraphics[width=0.90\textwidth]{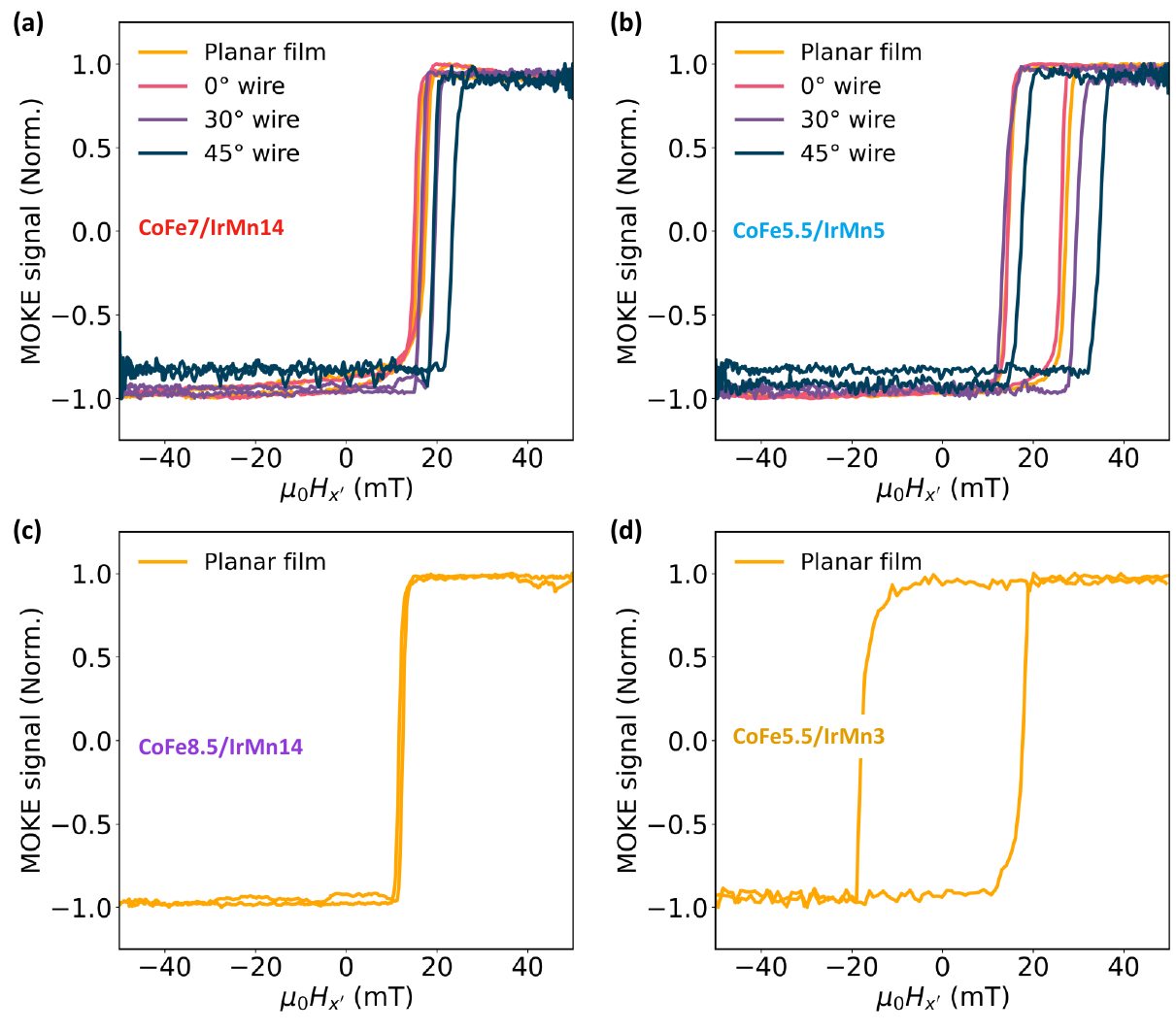}
    \caption*{\textbf{Fig. S1. Magnetization reversal in FM/AF bilayer systems.} 
    DF-MOKE hysteresis loops for \textbf{(A)} CoFe7/IrMn14,
    \textbf{(B)} CoFe5.5/IrMn5,
    \textbf{(C)} CoFe8.5/IrMn14 and
    \textbf{(D)} CoFe5.5/IrMn3.
    The magnetic field ($\mu_{0}H_{x'}$) is applied along the $x'$ direction (see Fig. 1a, for the coordinate systems).}
\end{figure}

\begin{figure*}[t]
    \centering
    \includegraphics[width=1.0\linewidth]{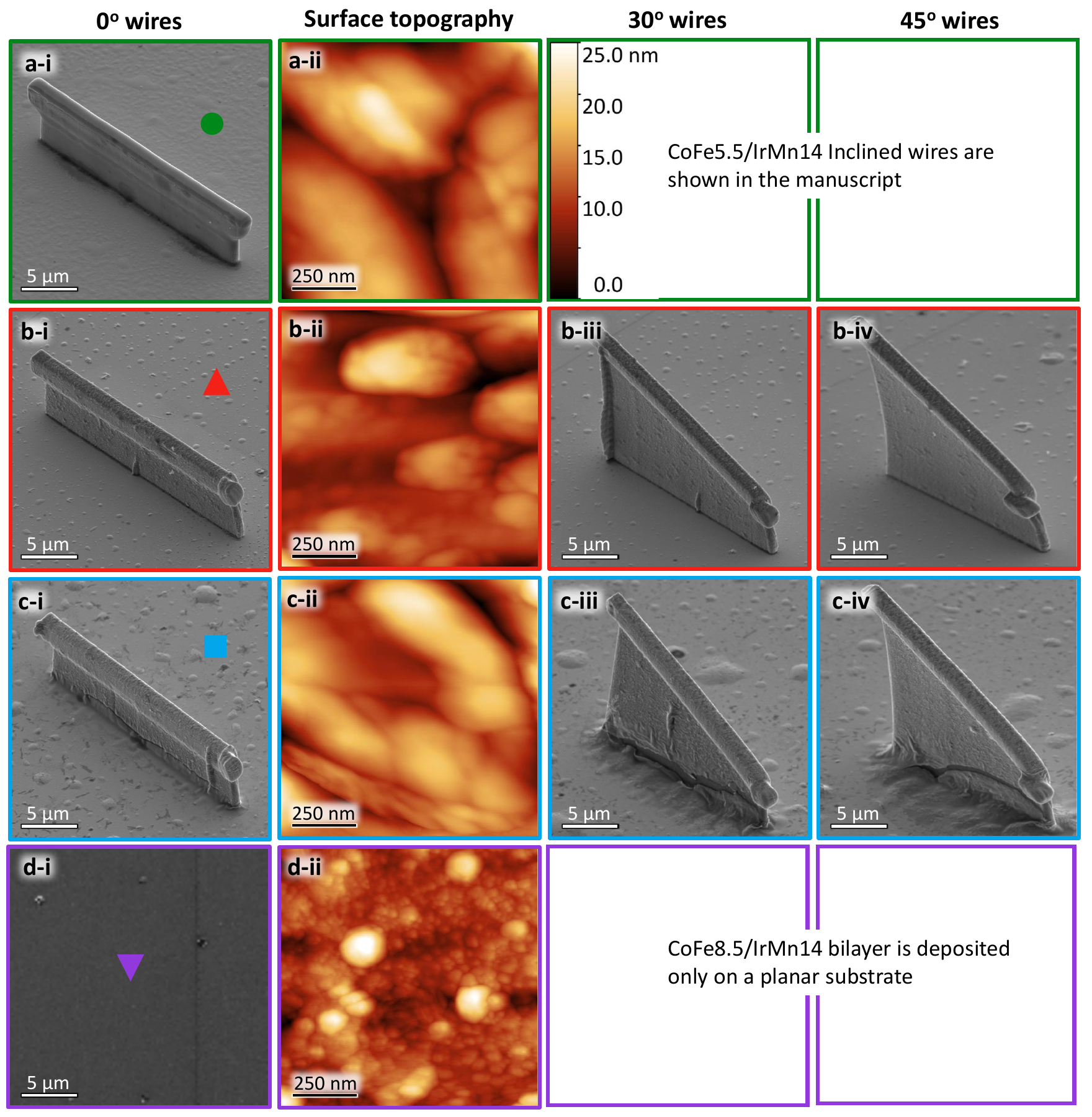}
   \caption*{\textbf{Fig. S2. AFM and SEM imaging of FM/AF systems.}
    AFM topography images of FM/AF planar films and SEM images of 3D FM/AF wires are shown for
    \textbf{(A)} CoFe5.5/IrMn14,
    \textbf{(B)} CoFe7/IrMn14,
    \textbf{(C)} CoFe5.5/IrMn5 and
    \textbf{(D)} CoFe8.5/IrMn14.}
    \label{fig:S2}
\end{figure*}

\twocolumngrid


\end{document}